\documentclass[english,preprint]{aastex}
\usepackage[T1]{fontenc}
\usepackage[latin1]{inputenc}

\makeatletter
\usepackage{babel}
\makeatother
\begin{document}

\title{The alignment of the magnetic field and collimated outflows in star
forming regions - the case of NGC 2071}

\author{Martin Houde \altaffilmark{1,2},\email{houde@ulu.submm.caltech.edu}
Thomas G. Phillips \altaffilmark{3}, Pierre Bastien\altaffilmark{2},
Ruisheng Peng\altaffilmark{1} \and Hiroshige Yoshida \altaffilmark{1}}

\affil{\altaffiltext{1}{Caltech Submillimeter Observatory, 111 Nowelo Street, Hilo, HI 96720}}

\affil{\altaffiltext{2}{Département de Physique, Université de Montréal, Montréal, Québec H3C 3J7, Canada}}

\affil{\altaffiltext{3}{California Institute of Technology, Pasadena, CA 91125}}

\begin{abstract}
The magnetic field is believed to play a crucial role in the process
of star formation. From the support it provides during the initial
collapse of molecular clouds to the creation of strong collimated
jets responsible for large mass losses, current theories predict its
importance in many different stages during the formation of stars.
Here we report on observational evidence which tests one aspect that
can be inferred from these theories: the alignment between the local
magnetic field and collimated bipolar outflows in such environments.
There is good evidence of an alignment in the case of NGC 2071.
\end{abstract}


\section{Introduction.}

In two previous papers \citep{Houde 2000a, Houde 2000b}, we presented
theory and observations which established the possibility of detecting
the presence of the magnetic field in molecular clouds through its
effects on the profile of molecular ion emission lines. This new effect
can be dramatic and can produce ion profiles with narrower line width
and significantly suppressed high velocity wings when compared to
the spectra of coexistent neutral molecular species. It is observed
in the weakly ionized plasmas of the dense interstellar medium provided
they are supported by turbulence or stellar outflows rather than thermally.
It also requires some degree of misalignment between the local mean
magnetic field and the neutral flows present in the gas. Differences
between the line profiles of the ion and neutral species disappear
if the flows and the magnetic field are aligned. 

This effect resembles in many ways the ambipolar diffusion phenomenon
which arises when one considers similar environments in a collisionally
dominated regime (i.e., magnetohydrodynamics). Under such circumstances,
if the small-scale spatial gradients of the magnetic field can be
neglected then gyromagnetic motions can also be ignored and the drift
speeds between the ions and the neutrals is found to be small \citep{Shu 1992}.
But as we have shown in our first paper on the subject \citep{Houde 2000a},
in the presence of neutral flows the magnetic forces quickly dominate
over collisional processes even in cases where the local field is
relatively weak (a few $\mu$G at a neutral density of $\sim10^{5}$
cm$^{-3}$). Gyromagnetic motions then become important, spatial variation
in the magnetic field on the scale of a gyro-radius arise and much
larger drift speeds are observed. 

As can be inferred from earlier comments, the detection of a magnetic
field is most easily established in cases where the differences between
the lines profiles of ions and neutrals are strongly accentuated.
However, under the assumption of the existence of a sufficiently large
magnetic field, a high degree of similarity in the spectra can potentially
tell us something about the alignment between the mean magnetic field
and the flows. This is mostly interesting for the study of objects
which show strong collimated bipolar outflows as it allows a comparison
between theory and observation. Indeed, current theoretical models
(see \citet{Bachiller 1996} for a review) predict the existence of
bipolar jets emanating from the poles of the protostars during certain
stages of star formation. One agent necessary for the presence of
these jets is believed to be the magnetic field to which they would
be aligned. 

While these jets can be observed at optical and/or radio frequencies,
molecular outflows, which appear to exist in irregular shells or in
the walls surrounding the cavities formed by the jets, are also detected
at millimeter and submillimeter wavelengths. The alignment between
these molecular outflows and the magnetic field has been observed
at optical wavelengths with measurements of the orientation of the
polarization vectors of the radiation emanating from background stars
located behind bipolar outflows originating from protostars \citep{Cohen 1984, Heiles 1987}
and, more recently, at millimeter or submillimeter wavelengths through
line and continuum polarization detections \citep{Girart 1999b}.
The technique we present here, which we argue also allows us to achieve
this, has the advantage of being very simple. It only requires comparisons
of the line profiles of coexisting neutral and ion molecular species,
in our case HCN and HCO$^{+}$, taken at the same positions in the
outflows.

\begin{figure}
\epsscale{0.7}\plotone{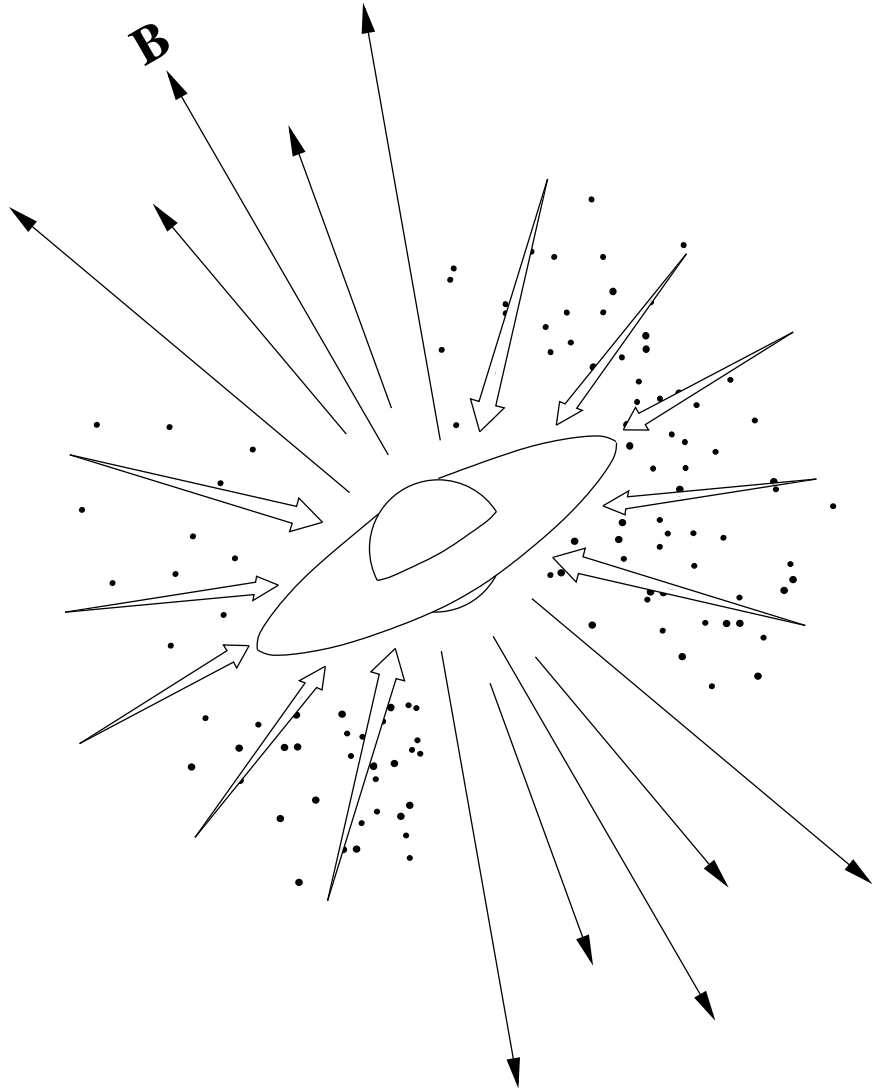}

\caption{\label{fig:geo}Example of a possible source geometry which exhibits
the symmetries enumerated in the text. The thick and thin arrows represent
infall and outflow, respectively, and the dots the ambient molecular
cloud.}
\end{figure}

To show how this can be done, we examine the following set of equations
\citep{Houde 2000b} which applies to cases where the region under
study has i) an azimuthal symmetry about the axis parallel to the
orientation of the mean magnetic field and ii) a reflection symmetry
across the plane perpendicular to this axis (see Figure \ref{fig:geo}).
We also assume that the flows in the region under study are linear,
the plasma is weakly ionized and all the collisions between ion and
neutral molecules are perfectly elastic:

\begin{eqnarray}
\left[\sigma_{obs}^{n}\right]^{2} & = & \sum_{k}C^{k}\left\langle \left[\mathbf{v}^{n}\right]^{k}\right\rangle ^{2}\left[\left\langle \cos\left(\theta^{k}\right)\right\rangle ^{2}\cos^{2}\left(\alpha\right)+\frac{1}{2}\left\langle \sin\left(\theta^{k}\right)\right\rangle ^{2}\sin^{2}\left(\alpha\right)\right]\label{eq:v2n}\\
\left[\sigma_{obs}^{i}\right]^{2} & \simeq & \sum_{k}C^{k}\left\langle \left[\mathbf{v}^{n}\right]^{k}\right\rangle ^{2}\left[\rule{0in}{5ex}\left\langle \cos\left(\theta^{k}\right)\right\rangle ^{2}\cos^{2}\left(\alpha\right)\right.\nonumber \\
 &  & \,\,\,\,\,\,\,\,\,\,\,\,\left.+\frac{\left\langle \sin\left(\theta^{k}\right)\right\rangle ^{2}}{\left[\frac{m_{i}}{\mu}-1\right]}\left[a\cos^{2}\left(\alpha\right)+\frac{g}{2}\sin^{2}\left(\alpha\right)\right]\right]\,.\label{eq:v2ion}\end{eqnarray}

The first equation gives an expression for the width (variance) of
the neutral line profile and the second another expression for that
of the ion. The summation runs over every flow present in any quadrant
of any plane which is perpendicular to the plane of reflection symmetry
and which also contains the axis of symmetry. $C^{k}$ is the weight
associated with the neutral flow $k$, which presumably scales with
the particle density (we assume ions and neutrals to exist in similar
proportions), its velocity is given by $\left[\mathbf{v}^{n}\right]^{k}$,
$\theta^{k}$ is the angle between its direction and that of the mean
magnetic field and $\alpha$ is the angle between the latter and the
line of sight to the observer. $a$ and $g$ are coefficients that
are weak functions of the ion mass $m_{i}$ and the reduced mass $\mu$.
If we choose the mean neutral molecular mass to be $2.3\, m_{\mbox{\scriptsize H}}$,
we get $a\simeq0.16$ and $g\simeq0.84$ for a wide range of ion masses.
We can easily verify from equations (\ref{eq:v2n}) and (\ref{eq:v2ion})
that, as was mentioned above, the ion and neutral species will have
similar line widths if the magnetic field and the flows are aligned
($\theta^{k}\simeq0$), irrespective of what the viewing angle $\alpha$
is. The difference between the line widths is maximized when the flows
are perpendicular to the field ($\theta^{k}\simeq\frac{\pi}{2}$). 

The difference between the two line widths is also a function of the
viewing angle $\alpha$. While we have $\sigma_{obs}^{n}\simeq\sigma_{obs}^{i}$
when the magnetic field is parallel to the line of sight ($\alpha=0$),
we get:

\[
\frac{\left[\sigma_{obs}^{i}\right]^{2}}{\left[\sigma_{obs}^{n}\right]^{2}}\simeq\frac{g}{\left[\frac{m_{i}}{\mu}-1\right]}\simeq0.07\]

\noindent for HCO$^{+}$ when $\alpha=\frac{\pi}{2}$ (with $\theta^{k}\neq0$
 for at least one value of $k$).

In general, the line narrowing and high velocity wing suppression
effects are seen in the core of molecular clouds for many ion species
(HCO$^{+}$, H$^{13}$CO$^{+}$, N$_{2}$H$^{+}$, HCS$^{+}$ and
H$_{3}$O$^{+}$ have been studied so far) whether the observed lines
are optically thin or not. Some notable examples are W3 IRS 5, OMC-1,
OMC-2 FIR 4 and OMC-3 MMS 6. We also showed, in our previous study
of a sample of eleven molecular clouds \citep{Houde 2000b}, that
the ratio of ion to neutral line width is likely to vary significantly
from one source to another, with the ion species always exhibiting
narrower line profiles. However, when averaged over the sample, it
is fairly close to what would be expected if the flows were randomly
oriented in relation to the direction of the mean magnetic field (as
can be calculated from equations (\ref{eq:v2n}) and (\ref{eq:v2ion})).
Perhaps this is a sign that there is no obvious propensity for the
alignment between the two in the cores of such high density molecular
clouds. But in at least one case of a well defined outflow, field
and flow alignment is indicated.

In the next section, we present observations of the NGC 2071 molecular
cloud where we argue that HCN and HCO$^{+}$ spectra provide evidence
for the alignment between the magnetic field and the collimated outflow.
This conclusion will be drawn from the remarkable similarities between
the profiles observed in the bipolar outflow for these two molecular
species. We should stress that this result, in itself, does not validate
the different theories that call for such an alignment. For all we
know NGC 2071 may just be a chance happening. Similar studies, like
the one presented here or others using different techniques, will
have to be done on a number of objects before a conclusion can be
reached.

In light of the above discussion concerning the significant difference
in line widths which we observed in molecular clouds, we infer that,
from a more general standpoint, the case studied in this paper is
rather unusual. An explanation for this might reside in the fact that
the aforementioned narrowing of the ion lines could be an indication
that the central structure of the clouds is more complex than the
simple, bipolar outflow that will be dealt with here. A given core
could harbor multiple flows and even, perhaps, a magnetic field which
may be controlled from a system external to it. For example, one could
conceive of the possible interactions between the fields and flows
emanating from different protostellar sources residing in the core.
Hopefully, possible future high-resolution (interferometric) observations
will shed some light on this issue.

\section{Observations.}

NGC 2071 is an optical reflection nebula located at approximately
390 pc in the L1630 molecular cloud of the Orion B region. This object
is known to exhibit a strong bipolar outflow which has been extensively
studied with many different molecular species (see \citet{Girart 1999a}
and references therein). Some of its main characteristics are its
relatively high degree of collimation, its large size and a favorable
viewing angle which all contribute to a relatively well understood
geometry, making this source a perfect candidate for the application
of our technique.

Our observations of NGC 2071 were made at the Caltech Submillimeter
Observatory (CSO) using the facility's 300-400 GHz receiver (beam
width $\simeq20\arcsec$). $^{12}$CO ($J\rightarrow3-2$) data was
obtained on 29 January 1997 and 5 February 1999 while HCN ($J\rightarrow4-3$)
and HCO$^{+}$ ($J\rightarrow4-3$) spectra were obtained on 29 October
1999 and 16 December 1999.

\begin{figure}
\notetoeditor{the four EPS files should appear in the figure fig:maps as shown with the commands below}

\resizebox*{16cm}{12cm}{\rotatebox{270}{\includegraphics{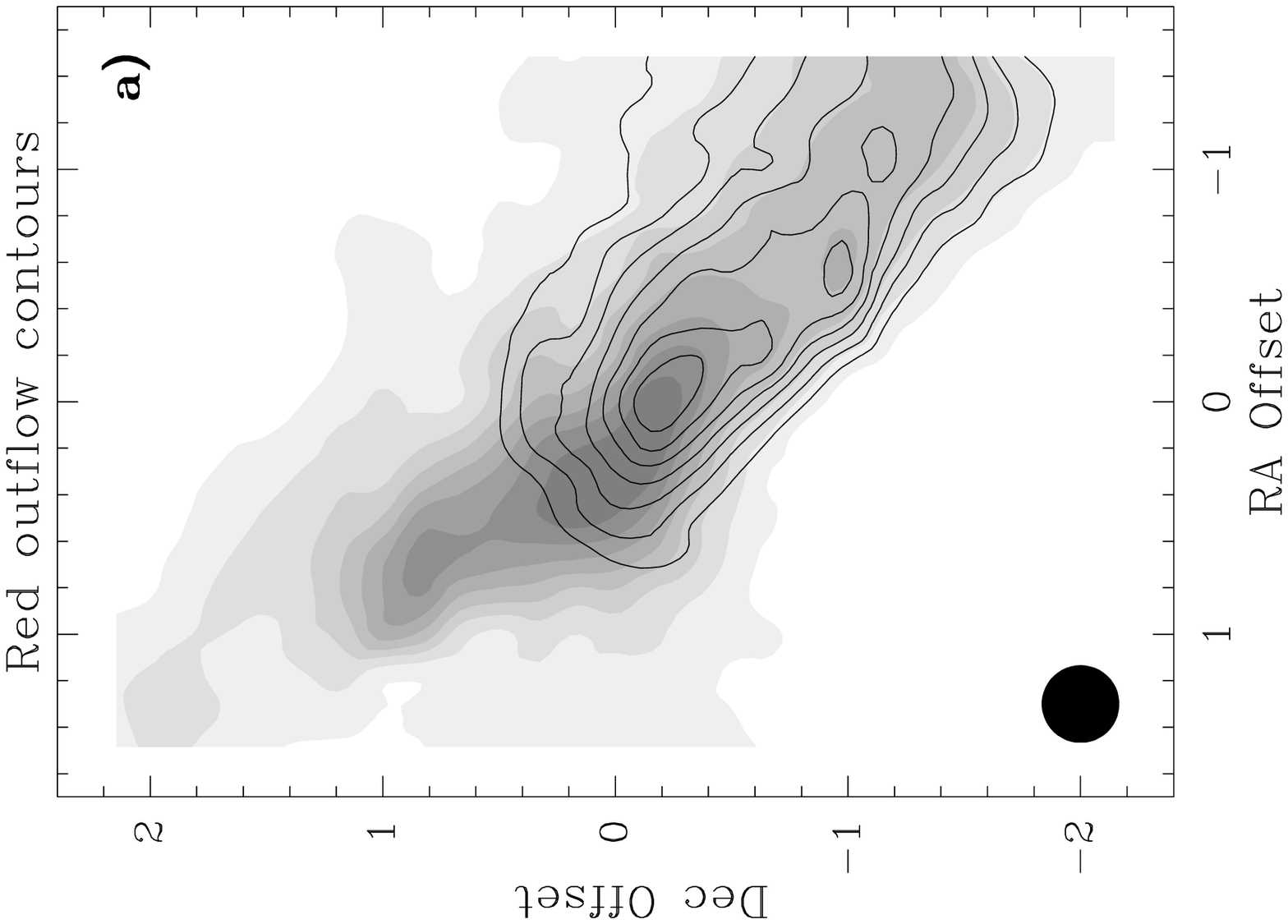}}\hspace{0.5cm}\rotatebox{270}{\includegraphics{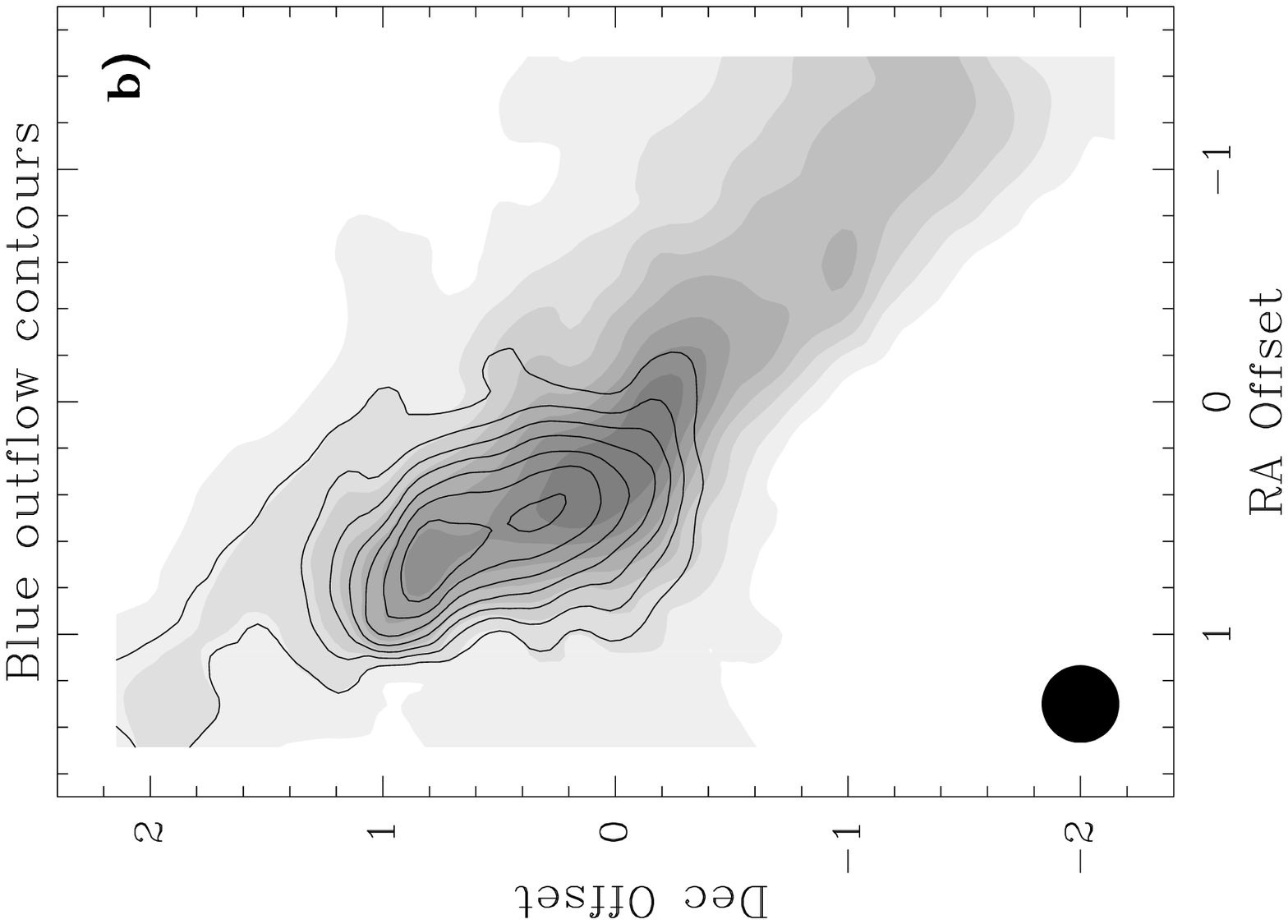}}}

\vspace{1cm}

\caption{\label{fig:maps}$^{12}$CO ($J\rightarrow3-2$) maps of NGC 2071.
Map a) shows contours for the integrated intensity (10 km/s $\leq v\leq$
50 km/s) of the red shifted lobe and map b) that of the blue shifted
lobe ($-50$ km/s $\leq v\leq$ 10 km/s). The underlying grey levels
delineate the total integrated intensity ($-50$ km/s $\leq v\leq$
50 km/s). The maps have a grid spacing of 10$\arcsec$. The size of
the telescope beam (FWHM$\simeq20\arcsec$) is shown in the lower
left corners. The center position of the maps is RA = $5^{\mathrm{h}}44^{\mathrm{m}}30\fs2 $,
DEC = $ 00\arcdeg20\arcmin42\farcs0 $ (J1950).}
\end{figure}

Figure \ref{fig:maps} shows two $^{12}$CO ($J\rightarrow3-2$) maps,
the contours highlight the integrated intensity of a) the red outflow
(10 km/s $\leq v\leq$ 50 km/s) and b) the blue outflow (-50 km/s
$\leq v\leq$ 10 km/s). The underlying grey scale levels delineate
the total integrated intensity (-50 km/s $\leq v\leq$ 50 km/s). 

\placefigure{fig:maps}

\begin{figure}
\notetoeditor{the four EPS files should appear in the figure fig:spectra1 as shown with the commands below}

\begin{center}\epsscale{0.9}\plottwo{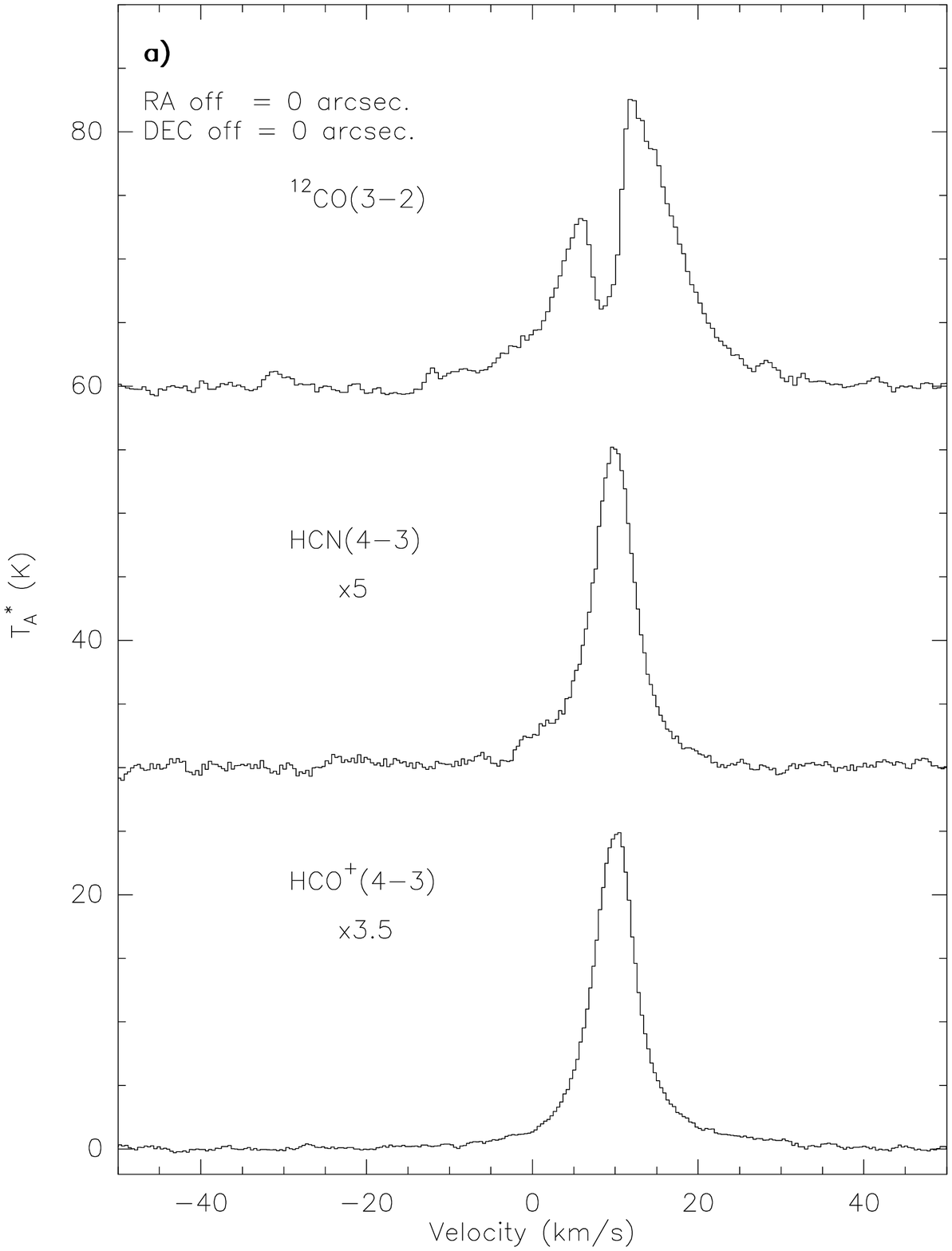}{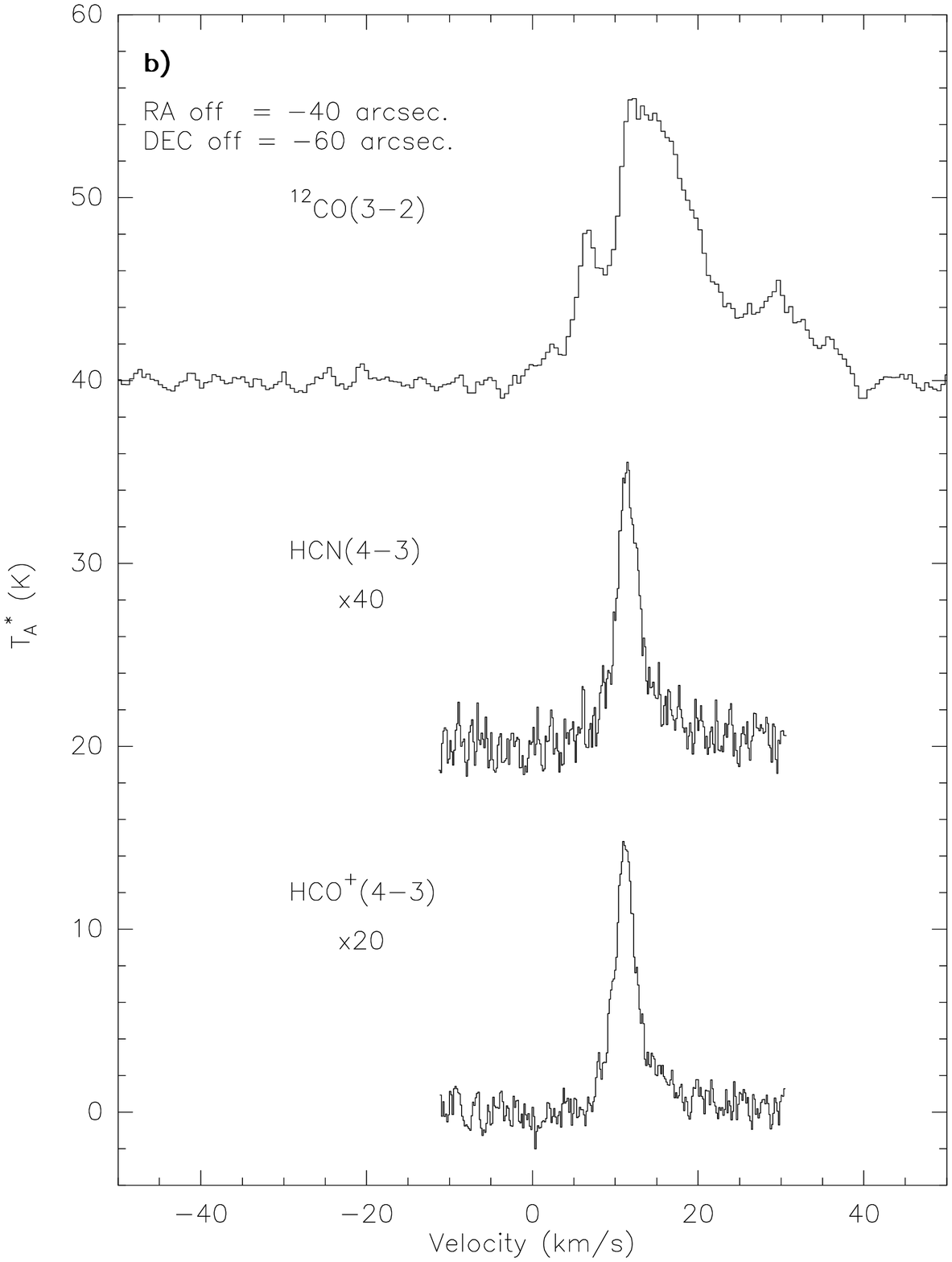}\end{center}

\begin{center}\epsscale{0.9}\plottwo{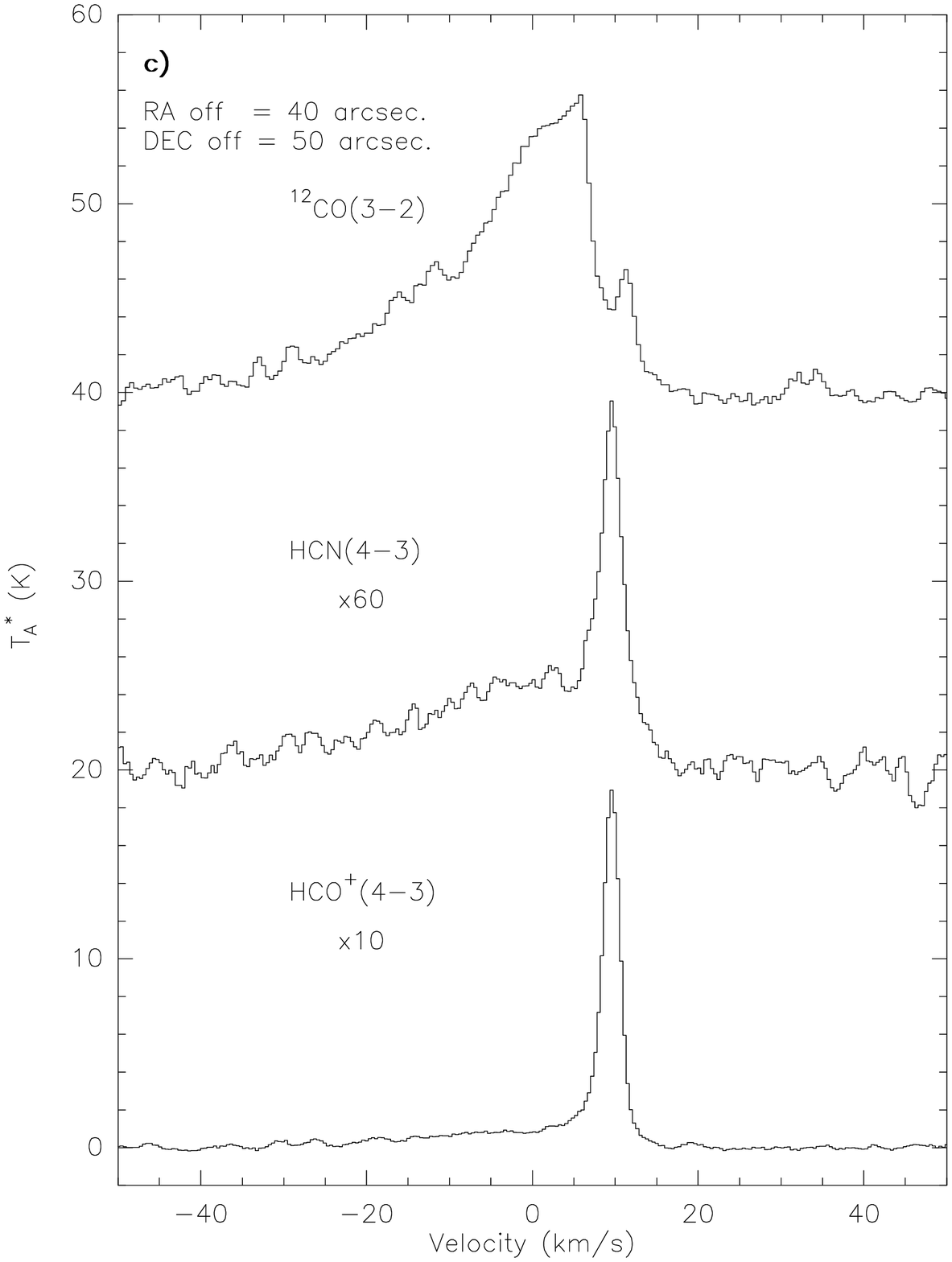}{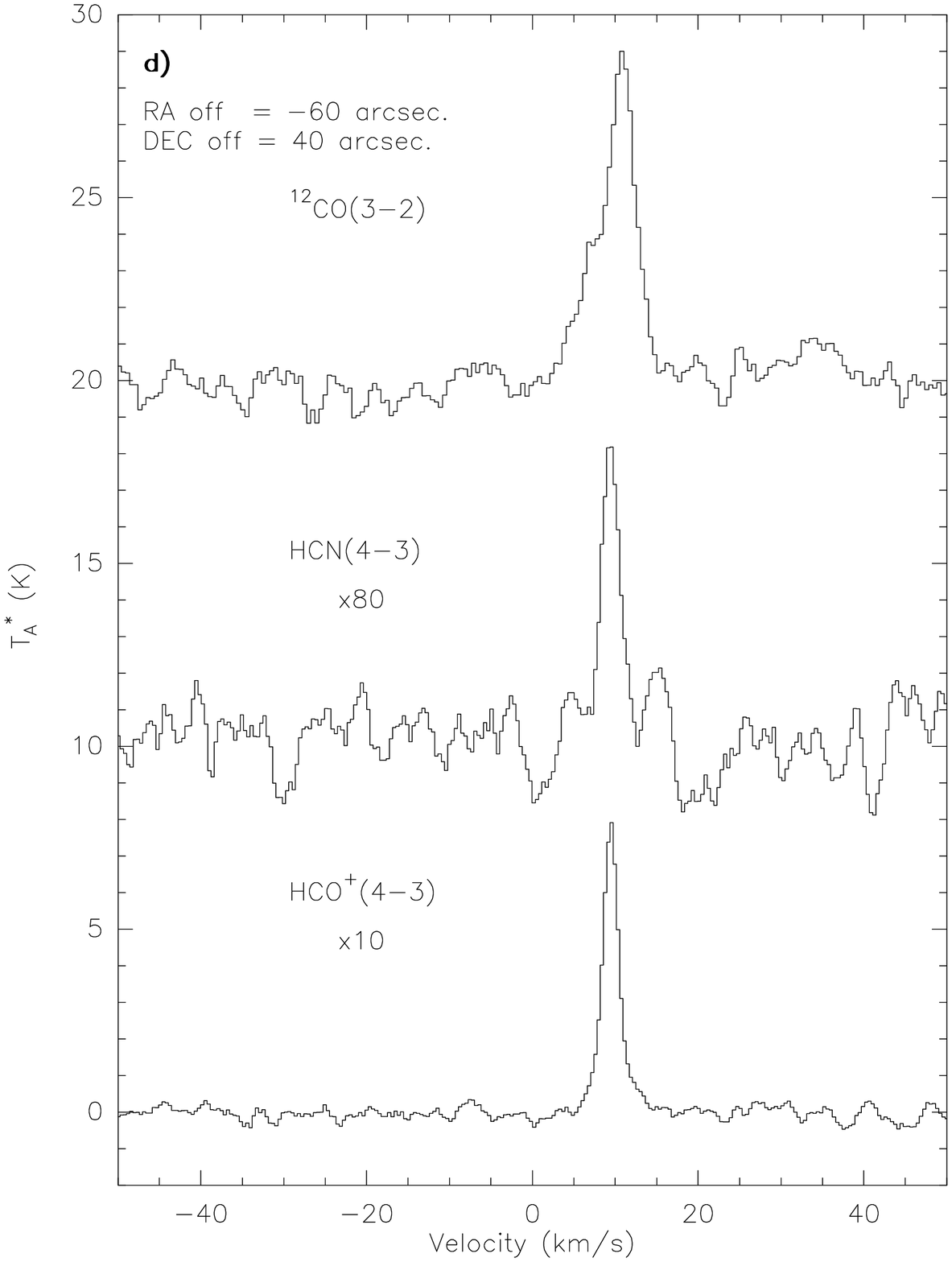}\end{center}

\caption{\label{fig:spectra1}$^{12}$CO ($J\rightarrow3-2$), HCN ($J\rightarrow4-3$)
and HCO$^{+}$ ($J\rightarrow4-3$) spectra in four different positions
in the NGC 2071 molecular cloud. Emission line profiles a) at the
center position ($\mbox{RA\, off.}=0\arcsec$, $\mbox{DEC\, off.}=0\arcsec$
in the maps shown in Figure \ref{fig:maps}), b) in the redshifted
outflow ($\mbox{RA\, off.}=-40\arcsec$, $\mbox{DEC\, off.}=-60\arcsec$),
c) in the blueshifted outflow ($\mbox{RA\, off.}=40\arcsec$, $\mbox{DEC\, off.}=50\arcsec$)
and d) away from the three previous positions ($\mbox{RA\, off.}=-60\arcsec$,
$\mbox{DEC\, off.}=40\arcsec$).}
\end{figure}

Figure \ref{fig:spectra1}a) shows three spectra ($^{12}$CO, HCN
and HCO$^{+}$) taken at the source's center position. The HCN and
HCO$^{+}$ spectra are very similar with their center velocity ($\approx9.5$
km/s) reasonably well aligned with the self-absorption feature of
the $^{12}$CO spectrum (presumably caused by and at the velocity
of the ambient quiescent cloud). The HCO$^{+}$ to HCN line widths
ratio was calculated to be 0.93 \citep{Houde 2000b}. Although this
position is not located in any of the two outflows, this number suggests
that perhaps we have a good alignment between the mean magnetic field
and the outflow. To verify this, we have obtained deep integration
spectra in one position for each outflow.

Figure \ref{fig:spectra1}b) shows spectra of the same molecular species
taken in the redshifted outflow ($\mbox{RA\, off.}=-40\arcsec$, $\mbox{DEC\, off.}=-60\arcsec$).
The $^{12}$CO spectrum shows an extended high velocity wing that
goes as far as 40 km/s. The HCN and HCO$^{+}$ spectra are, of course,
much weaker at this position than in the core but also exhibit redshifted
velocity wings which indicate that they are taking part in the outflow
motion. At first look, these two spectra show similar profiles, reinforcing
the alignment hypothesis between the magnetic field and the collimated
outflow. Note that the velocity at peak intensity ($\approx11.2$
km/s) differs somewhat from that of the quiescent part of the cloud
($\approx9.5$ km/s). Figure \ref{fig:spectra1}c) shows a similar
set of spectra taken in the blueshifted outflow ($\mbox{RA\, off.}=40\arcsec$,
$\mbox{DEC\, off.}=50\arcsec$). The $^{12}$CO spectrum again exhibits
a strong high velocity wing that extends to approximately $-50$ km/s.
This time, the HCN and HCO$^{+}$ line profiles are noticeably different.
But both spectrum have a velocity wing that extends as far as that
of $^{12}$CO while also showing a narrow peak at the velocity of
the ambient cloud. This component can be discarded when comparing
the wing profiles. This can be ascertained with the help of Figure
\ref{fig:spectra1}d) where we present a set of spectra taken at a
position ($\mbox{RA\, off.}=-60\arcsec$, $\mbox{DEC\, off.}=40\arcsec$)
situated at approximately the same distance from the core as those
of Figure \ref{fig:spectra1}b) and c) but in a direction roughly
perpendicular to the axis of the outflow. At this position, only this
component is present in the spectra and it is natural to assume that
it is the result of emissions emanating from the quiescent part of
the cloud and not from the outflowing gas.

\placefigure{fig:spectra1}

\section{Alignment of the magnetic field and the collimated outflows.}

\begin{figure}
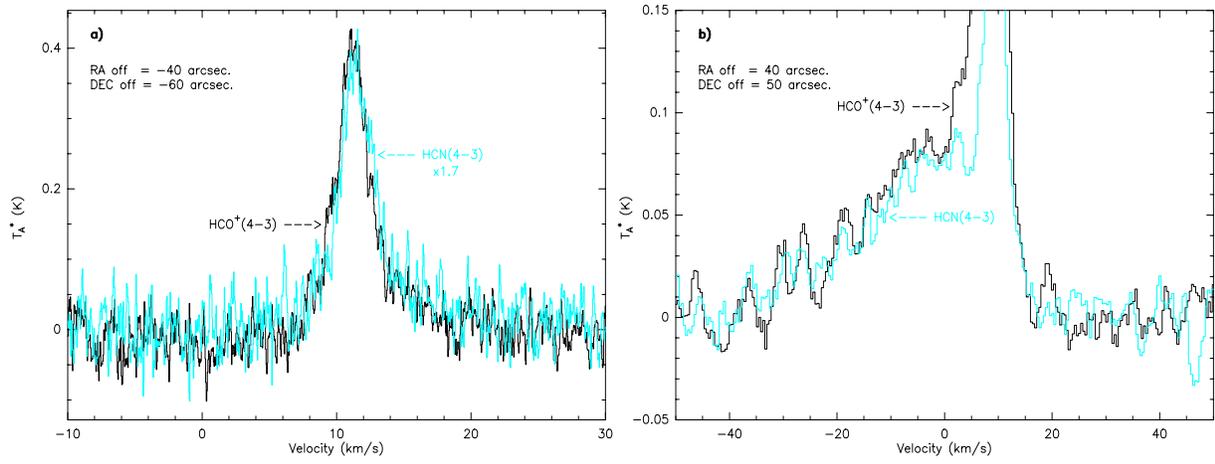

\notetoeditor{the four EPS files should appear in the figure fig:spectra2 as shown with the commands below}

\resizebox*{16cm}{6cm}{\rotatebox{270}{\includegraphics{f4a_3.eps.rgb}}\hspace{0.2cm}\rotatebox{270}{\includegraphics{f4b_3.eps.rgb}}}

\vspace{1cm}

\caption{\label{fig:spectra2}Superposition of the HCN ($J\rightarrow4-3$)
and HCO$^{+}$ ($J\rightarrow4-3$) line profiles in the a) redshifted
and b) blueshifted outflows. The profiles are remarkably similar for
velocities different from that of the ambient cloud ($v\simeq9.5$
km/s). Note that the two figures have different velocity scales.}
\end{figure}

As we mentioned earlier, according to the material we presented in
our two previous papers \citep{Houde 2000a, Houde 2000b}, the line
profiles of coexistent ion and neutral molecular species should show
no significant differences in cases where there is a good alignment
between the mean magnetic field and the flows present in a given region
of a molecular cloud. In Figure \ref{fig:spectra2}, we have superposed
the HCN and HCO$^{+}$ spectra obtained in the red and blueshifted
outflows presented in Figure \ref{fig:spectra1}b) and c). We have
also scaled the range of the vertical axis in order to better show
the details of the different spectra and avoid any unwanted confusion
that could be brought by the components of the quiescent part of the
cloud. In both cases, the similarity between the HCN and HCO$^{+}$
spectra is striking. We are led to the conclusion that there is strong
evidence for a high degree of alignment between the magnetic field
and the bipolar outflow of NGC 2071. We do not mean to imply that
this validates the different theories that call for such an alignment,
for all we know NGC 2071 may just be a chance happening. Similar studies
like this one (or using other techniques) will have to be done on
other objects before a conclusion can be reached.

Evidently, there is always the alternative that there is no magnetic
field in the outflows to impede the motion of the ions (a comparison
of optically thin lines (H$^{13}$CN and H$^{13}$CO$^{+}$) which
probe deeper in the source's center, presented in \citet{Houde 2000b},
gives an ion to neutral line width ratio of 0.64 and, therefore, establishes
the presence of a magnetic field in the core). Because of this we
cannot be absolutely certain of the alignment. But as \citet{Cohen 1984}
and \citet{Girart 1999b} have shown, a magnetic field does exist
in the outflow of many objects. It is therefore likely that the same
is true for outflowing sources in general.

\placefigure{fig:spectra2}

\subsection{The driving agent behind the outflows.}

The previous discussion and the conclusion we reached are partly based
on the assumption, and likely hypothesis, that jets are the primary
driving agent of molecular outflows. In this scenario, outflows with
standard high velocities would consist of ambient cloud gas which
has been swept up by underlying winds. Whether these winds are composed
of the (ionized) jets alone or in combination with a significant neutral
component is uncertain, but there is evidence that entrainment of
the ambient gas by bow shocks or turbulent mixing on the sides of
the jets can be achieved \citep{Bachiller 1996}.

Our interpretation of the data might change if the driving agent were
of a different nature. In fact, our analysis rests on the assumption
that neutral flows are present, working on the ion species and trying
to bring them along in the flows via collisions. Something of this
sort is needed to explain the difference in the line profiles between
coexistent ions and neutrals observed in the core of molecular clouds
(i.e., the larger widths of the neutral spectra \citep{Houde 2000a, Houde 2000b}).
Our conclusions might not apply if, on the other hand, the situation
were reversed and the ions were driven through some magnetohydrodynamical
process and working on the neutrals. But we would then expect the
ions to have spectra which would be at least as wide as those of the
neutrals and sometimes broader, depending on the degree of coupling
between the two species.

\section{Conclusion.}

Using some of the material presented in our two previous papers \citep{Houde 2000a, Houde 2000b},
we have shown how it is possible to find out if the mean magnetic
field and collimated outflows are aligned with each other in star
formation regions. We have applied this technique to data obtained
at the CSO for the NGC 2071 molecular cloud and concluded that there
is strong evidence that there exists a high degree of alignment in
this case. This technique has the advantage of being very simple since
it only requires a comparison between the line profiles of two coexistent
molecular species (in our case HCN and HCO$^{+}$).

\acknowledgements{M. Houde's work was done in part with the assistance of grants from
FCAR and the Département de Physique of the Université de Montréal.
The Caltech Submillimeter Observatory is funded by the NSF through
contract AST 9615025. }


\end{document}